\documentclass[aps,prd,eqsecnum,preprint,tightenlines,nofootinbib]{revtex4-1}
\usepackage{graphicx,latexsym}

\newcommand{\bea}{\begin{eqnarray}}
\newcommand{\eea}{\end{eqnarray}}
\newcommand{\be}{\begin{equation}}
\newcommand{\ee}{\end{equation}}

\newcommand{\Pminus}{{\cal P}^-}

\newcommand{\pp}{p^{\prime +}}

\begin{document}

\title{Light-front $\phi^4_{1+1}$ theory \\
using a many-boson symmetric-polynomial basis%
\footnote{Based on a talk contributed to the
Lightcone 2015 workshop, Frascati, Italy, 
September 21-25, 2015.}
}

\author{S.S. Chabysheva}
\affiliation{Department of Physics and Astronomy\\
University of Minnesota-Duluth \\
Duluth, Minnesota 55812}

\date{\today}

\begin{abstract}
We extend earlier work on fully symmetric polynomials
for three-boson wave functions to arbitrarily many
bosons and apply these to a light-front analysis
of the low-mass eigenstates of $\phi^4$ theory in
1+1 dimensions.  The basis-function approach allows
the resolution in each Fock sector to be independently
optimized, which can be more efficient than the preset
discrete Fock states in DLCQ.  We obtain an estimate
of the critical coupling for symmetry breaking
in the positive mass-squared case.
\end{abstract}

\maketitle

\section{Introduction} \label{sec:intro}

Our objective is to use newly developed multivariate
polynomials~\cite{GenSymPolys} as a basis set for the computation of
the odd and even-parity massive eigenstates of light-front
$\phi_{1+1}^4$ theory~\cite{RozowskyThorn,Varyetal}.  The eigenvalue problem is
solved in the form of a truncated Fock-state expansion, with
each Fock wave function expanded in the polynomials.
This allows separate tuning of the resolution in each
Fock sector, which should be more efficient than a 
discrete light-cone quantization (DLCQ)~\cite{PauliBrodsky,DLCQreview}
calculation~\cite{VaryHari}.
We can then study convergence with respect to the Fock-space 
truncation and compare results with 
those obtained with the light-front coupled-cluster (LFCC)
method~\cite{LFCC,LFCCphi4}.
We also estimate the value of the critical coupling
for symmetry breaking.

The Lagrangian for two-dimensional $\phi^4$ theory is
${\cal L}=\frac12(\partial_\mu\phi)^2-\frac12\mu^2\phi^2-\frac{\lambda}{4!}\phi^4$,
where $\mu$ is the mass of the boson and $\lambda$ is the coupling constant.
The light-front Hamiltonian density is
${\cal H}=\frac12 \mu^2 \phi^2+\frac{\lambda}{4!}\phi^4$.
The mode expansion for the field at zero light-front time is
\be \label{eq:mode}
\phi=\int \frac{dp^+}{\sqrt{4\pi p^+}}
   \left\{ a(p^+)e^{-ip^+x^-/2} + a^\dagger(p^+)e^{ip^+x^-/2}\right\},
\ee
with the modes quantized such that
$[a(p^+),a^\dagger(\pp)]=\delta(p^+-\pp)$.
The light-front Hamiltonian is 
$\Pminus=\Pminus_{11}+\Pminus_{13}+\Pminus_{31}+\Pminus_{22}$,
with
\bea \label{eq:Pminus11}
\Pminus_{11}&=&\int dp^+ \frac{\mu^2}{p^+} a^\dagger(p^+)a(p^+),  \\
\label{eq:Pminus13}
\Pminus_{13}&=&\frac{\lambda}{6}\int \frac{dp_1^+dp_2^+dp_3^+}
                              {4\pi \sqrt{p_1^+p_2^+p_3^+(p_1^++p_2^++p_3^+)}} 
     a^\dagger(p_1^++p_2^++p_3^+)a(p_1^+)a(p_2^+)a(p_3^+), \\
\label{eq:Pminus31}
\Pminus_{31}&=&\frac{\lambda}{6}\int \frac{dp_1^+dp_2^+dp_3^+}
                              {4\pi \sqrt{p_1^+p_2^+p_3^+(p_1^++p_2^++p_3^+)}} 
      a^\dagger(p_1^+)a^\dagger(p_2^+)a^\dagger(p_3^+)a(p_1^++p_2^++p_3^+), \\
\label{eq:Pminus22}
\Pminus_{22}&=&\frac{\lambda}{4}\int\frac{dp_1^+ dp_2^+}{4\pi\sqrt{p_1^+p_2^+}}
       \int\frac{dp_1^{\prime +}dp_2^{\prime +}}{\sqrt{p_1^{\prime +} p_2^{\prime +}}} 
       \delta(p_1^+ + p_2^+-p_1^{\prime +}-p_2^{\prime +})
  a^\dagger(p_1^+) a^\dagger(p_2^+) a(p_1^{\prime +}) a(p_2^{\prime +}) .
\eea

The eigenstate with momentum $P^+$ is expanded as
\be \label{eq:FSexpansion}
|\psi(P^+)\rangle=\sum_m (P^+)^{\frac{m-1}{2}}\int\prod_i^m dy_i 
       \delta(1-\sum_i^m y_i)\psi_m(y_i)\frac{1}{\sqrt{m!}}\prod_{i=1}^m a^\dagger(y_iP^+)|0\rangle.
\ee
The sum over $m$ is restricted to odd or even numbers.
The Hamiltonian does not mix the two cases, and we solve
for the lowest eigenstate in each case.

With use of the Fock-state expansion, the light-front Hamiltonian 
eigenvalue problem
${\cal P}^-|\psi(P)\rangle=\frac{M^2}{P}|\psi(P)\rangle$
becomes
\bea \label{eq:coupledsystem}
\lefteqn{\frac{m}{y_1}\psi_m(y_i)
+\frac{g}{4}\frac{m(m-1)}{\sqrt{y_1y_2}}
        \int\frac{dx_1 dx_2 }{\sqrt{x_1 x_2}}\delta(y_1+y_2-x_1-x_2) \psi_m(x_1,x_2,y_3,\ldots,y_m)}&& 
        \nonumber \\
& +\frac{g}{6}m\sqrt{(m+2)(m+1)}\int \frac{dx_1 dx_2 dx_3}{\sqrt{y_1 x_1 x_2 x_3}}
        \delta(y_1-x_1-x_2-x_3)\psi_{m+2}(x_1,x_2,x_3,y_2,\ldots,y_m) \nonumber \\
& +\frac{g}{6}\frac{(m-2)\sqrt{m(m-1)}}{\sqrt{y_1y_2y_3(y_1+y_2+y_3)}}
          \psi_{m-2}(y_1+y_2+y_3,y_4,\ldots,y_m)=\frac{M^2}{\mu^2}\psi_m(y_i).
\eea
Here $g=\lambda/4\pi\mu^2$ is a dimensionless coupling.  It is to this system
of equations that we apply our basis-function expansion for each Fock-state
wave function, to convert the coupled integral equations to a matrix 
eigenvalue problem.

\section{Solution of the integral equations} \label{sec:Solution}

We solve this coupled system by first truncating the Fock-state
expansion at some maximum number of constituents and then
expanding each wave function in
a basis of symmetric multivariate polynomials $P_{ni}^{(m)}$
\be \label{eq:expansion}
\psi_m(y_i)=\sqrt{\prod_i y_i}\sum_{ni} c_{ni}^{(m)} P_{ni}^{(m)}(y_1,\ldots,y_m).
\ee
The $P_{ni}^{(m)}$ are of order $n$ and 
fully symmetric with respect to interchange of momenta~\cite{GenSymPolys}.
The subscript $i$ differentiates the various possibilities at a 
given order $n$.  For $m=2$ constituents there is only one possibility
at each order, but for $m>2$ there can be more than one.
For example, for three constituents there are two sixth-order
polynomials, $P^{(3)}_{61}=(y_1y_2y_3)^2$ and $P^{(3)}_{62}=(y_1y_2+y_1y_3+y_2y_3)^3$.

The number of linearly independent polynomials of a given order is restricted
by the constraint of momentum conservation, $\sum_i y_i=1$.  For example,
$P^{(3)}_{2}=y_1y_2+y_1y_3+y_2y_3$ is equivalent to $y_1^2+y_2^2+y_3^2$, 
up to a constant, when $y_3$ is replaced by $1-y_1-y_2$.

The linearly independent symmetric polynomials can be written as
products of powers of simpler polynomials, in the form
$P_{ni}^{(N)}=C_2^{n_2} C_3^{n_3}\cdots C_N^{n_N}$,
with the powers restricted by $n=\sum_j j n_j$.
Each different
way of decomposing $n$ into a sum of integers greater than 1
yields a different polynomial.
The $C_m$ are sums of simple monomials $\prod_j^N y_j^{m_j}$,
where $m_j$ is 0 or 1 and $\sum_j^N m_j=m$.  The sum over
the monomials ranges over all
possible choices for the $m_j$, making each $C_m$ fully symmetric.

As examples of the $C_m$, consider the general case of 
$N$ longitudinal momentum variables.  Then $C_2$ is 
just $\sum_j^N \left(y_j\sum_{k>j}^N y_k\right)$;
$C_{N-1}$ is $\sum_j^N \prod_{k\neq j} y_k$; and
$C_N$ is $y_1y_2\cdots y_N$.  In particular, for N=3,
$C_2=y_1y_2+y_1y_3+y_2y_3$ and $C_3=y_1y_2y_3$.
The first-order polynomial
$C_1=\sum_j y_j$ does not appear because the momentum constraint reduces
it to a constant.

Projection of the coupled system onto the basis functions
yields the matrix equations
\be \label{eq:matrixequations}
\sum_{n'i'}\left[T^{(m)}_{ni,n'i'}+g V^{(m,m)}_{ni,n'i'}\right]c^{(m)}_{n'i'}
   +g \sum_{n'i'} V^{(m,m+2)}_{ni,n'i'} c^{(m+1)}_{n'i'}
   +g \sum_{n'i'} V^{(m,m-2)}_{ni,n'i'} c^{(m-1)}_{n'i'}
   =\frac{M^2}{\mu^2}\sum_{n'i'}B^{(m)}_{ni,n'i'}c_{n'i'}^{(m)},
\ee
with $T^{(m)}$ the kinetic-energy matrix in the $m$th Fock sector
\be
T^{(m)}_{ni,n'i'}=m\int\left(\prod_j dy_j \right)\delta(1-\sum_j y_j)\left(\prod_{j=2}^m y_j\right) P_{ni}^{(m)}(y_j)P_{n'i'}^{(m)}(y_j),
\ee
$B^{(m)}$ the basis-function overlap matrix
\be
B^{(m)}_{ni,n'i'}=\int\left(\prod_j dy_j \right)\delta(1-\sum_j y_j)\left(\prod_j^m y_j\right) P_{ni}^{(m)}(y_j)P_{n'i'}^{(m)}(y_j),
\ee
and $V^{(m,m')}$ the potential-energy matrices
\bea
V^{(m,m)}_{ni,n'i'}&=&\frac{g}{4}m(m-1)\int\left(\prod_j dy_j\right) \delta(1-\sum_j y_j \\
  && \times  \int dx_1 dx_2 \delta(y_1+y_2-x_1-x_2)
      \left(\prod_{j=3}^m y_j\right) P_{ni}^{(m)}(y_j)P_{n'i'}^{(m)}(x_1,x_2,y_3,\ldots,y_m),
      \nonumber \\
V^{(m,m+2)}_{ni,n'i'}&=&\frac{g}{6}m\sqrt{(m+2)(m+1)}\int\left(\prod_j dy_j\right) \delta(1-\sum_j y_j) \\
  && \times 
    \int dx_1 dx_2 dx_3 \delta(y_1-x_1-x_2-x_3)
      \left(\prod_{j=2}^m y_j\right) P_{ni}^{(m)}(y_j)P_{n'i'}^{(m+2)}(x_1,x_2,x_3,y_2,\ldots,y_m),
      \nonumber \\
V^{(m,m-2)}_{ni,n'i'}&=&\frac{g}{6}(m-2)\sqrt{m(m-1)}\int\left(\prod_j dy_j\right)
        \delta(1-\sum_j y_j) \\
  && \times 
      \left(\prod_{j=4}^m y_j \right)P_{ni}^{(m)}(y_j)P_{n'i'}^{(m-2)}(y_1+y_2+y_3,y_4,\ldots,y_m).
      \nonumber
\eea
All of the integrals can be done analytically in terms of a generalized beta function.

We now have a generalized eigenvalue problem of the form $H\vec c=(M^2/\mu^2) B\vec c$.
A standard approach would be to factorize $B$ and convert the original problem
to an ordinary eigenvalue problem.  However, factorization can fail in practice 
due to round-off errors in the implicit orthogonalization of the basis.  Round-off
errors also plague an explicit orthogonalization.  A reliable factorization is a 
singular-value decomposition $B=UDU^T$, where the columns of the matrix $U$ 
are the eigenvectors of $B$ and $D$ is a diagonal matrix of the eigenvalues of $B$.
We then solve $H'\vec c^{\,\prime}=(M^2/\mu^2) \vec c^{\,\prime}$, with
$H'=D^{-1/2}U^T HUD^{-1/2}$ and $\vec c^{\,\prime}=D^{1/2}U^T\vec c$.
Results for particular truncations of the polynomial basis are extrapolated
to an infinite basis size in each Fock sector.

\section{Results} \label{sec:Results}

\begin{figure}
\centering
\begin{tabular}{cc}
\includegraphics[width=7.5cm]{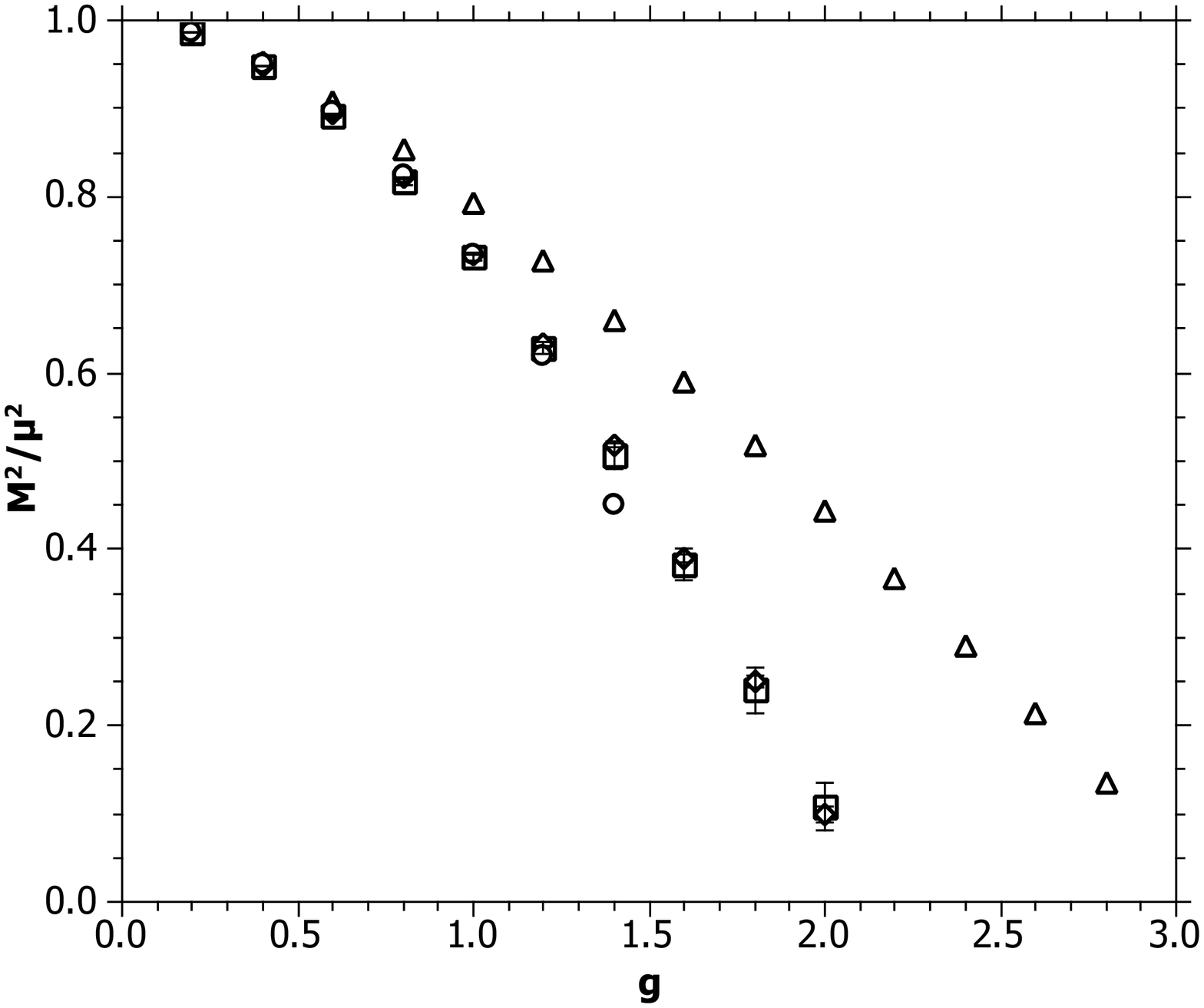} &
\includegraphics[width=7.5cm]{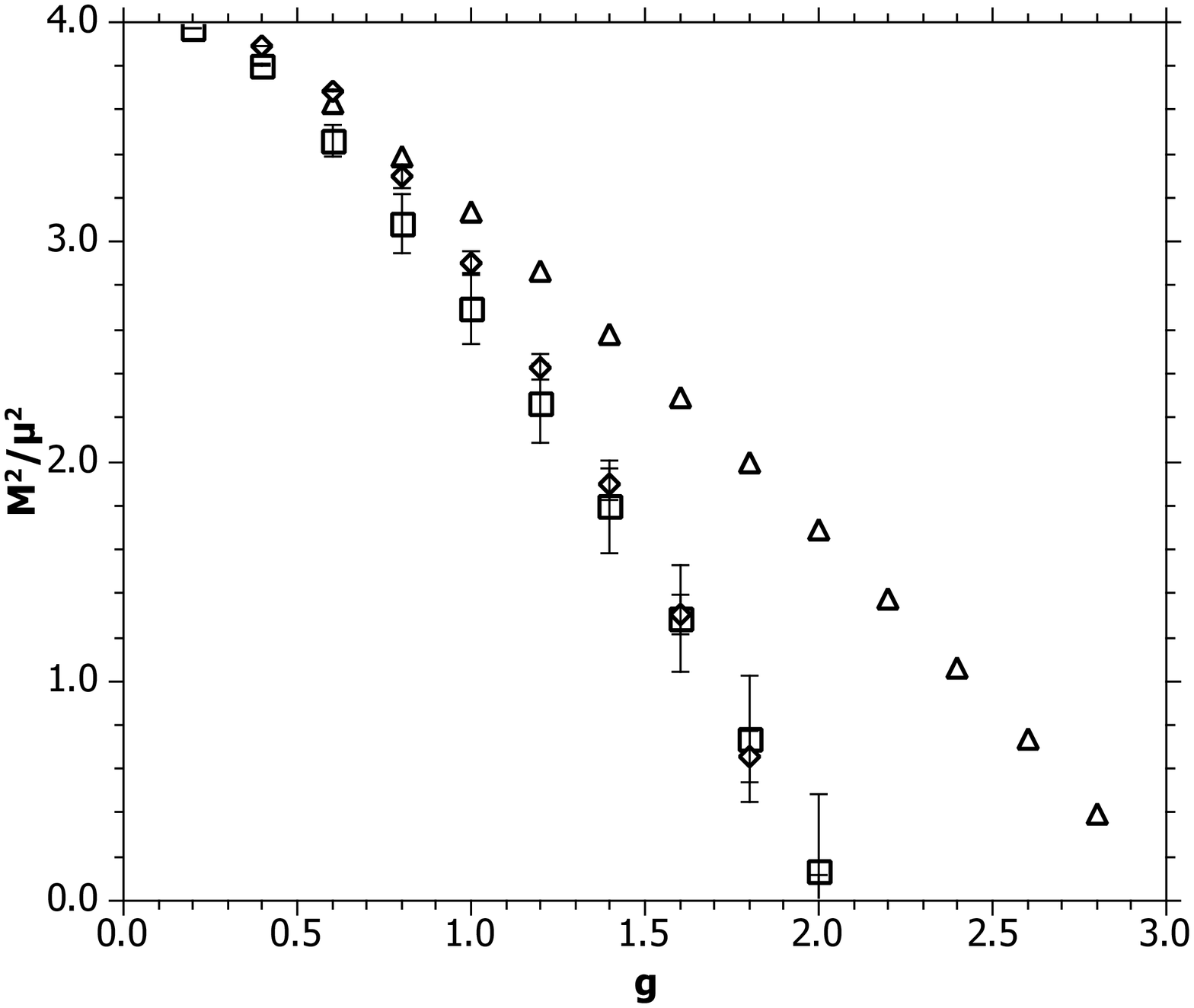} \\
(a) & (b) 
\end{tabular}
\caption{Mass squared vs coupling strength for an (a) odd
and (b) even number of constituents.  The different Fock-space
truncations in (a) are the three-body (triangles), five-body
(squares), and seven-body (diamonds) Fock sectors.  Results
for the LFCC method (circles) are also included.
In (b) the different truncations are the four-body (triangles),
six-body (squares), and eight-body (diamonds) Fock sectors.
Error bars are determined by the fits to extrapolation in
the polynomial basis size.}
\label{fig:extrap}
\end{figure}

The results for the odd and even cases are shown in Fig.~\ref{fig:extrap},
where $M^2$ is plotted in units of the bare mass squared $\mu^2$ for a
range of dimensionless coupling strengths $g$.  The convergence with
respect to Fock sector truncation is easily seen to be rapid, with the
last two truncations yielding identical results to within errors in
each case.  We also compare the lowest order LFCC results~\cite{LFCCphi4} for the 
odd case and find that these are quite consistent with the converged
Fock-space calculation, even though the LFCC calculation involves 
only a three-body function.

\begin{figure}
\centering
\includegraphics[width=10cm]{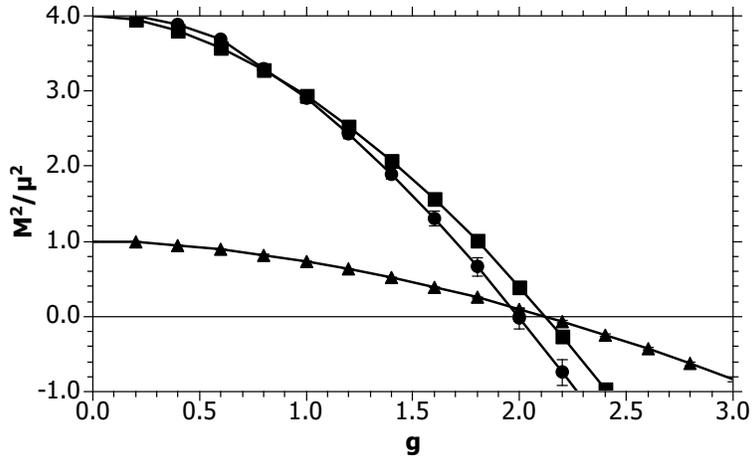}
\caption{Mass squared vs coupling strength, as used to estimate the critical
coupling.  The odd (triangles) and even (circles) cases are represented 
by the seven-body and eight-body truncations, respectively.  Points at
twice the mass of the odd case (squares) are also plotted.}
\label{fig:critcoup2}
\end{figure}

For both the odd and even cases, the mass crosses zero at
a finite value of the coupling, with the massive eigenstate
becoming degenerate with the Fock vacuum.  We interpret this 
as the appearance of symmetry breaking and extract a
value of the critical coupling, with use of the results
as plotted in Fig.~\ref{fig:critcoup2}.  The odd and even
cases cross zero at nearly the same value.  As a check,
we plot points for four times the $M^2$ values in the odd
case; in an exact calculation these should coincide with
the even values, and here they are close, consistent with
the errors in each.  From the
figure we estimate the critical coupling to be 
$g_c=2.1\pm0.05$.

Comparison with other calculations of the critical coupling,
as summarized in Table~\ref{tab:critcoup}, can be readily made.
However, the values compiled by Rychkov and Vitale~\cite{RychkovVitale}
are normalized in a slightly different manner; the translation
from our $g$ to theirs is $\bar{g}=\frac{\pi}{6}g$.
Clearly there is a systematic difference between equal-time and light-front values.
However, this is consistent with the expectation that the renormalization of the 
mass $\mu$ is different in the two quantizations~\cite{Burkardt}.
Calculations to evaluate this difference quantitatively are underway.
\begin{table}
\caption{Comparison of values for the critical coupling,
with $\bar{g}=\frac{\pi}{6}g$, as compiled in Ref.~\cite{RychkovVitale}.
The first two use light-front quantization; the remainder use equal-time
quantization.}
\centering
\label{tab:critcoup} 
\begin{tabular}{lll}
\hline\hline
Method &  $\bar{g}_c$ & Reported by  \\
\hline
LF symmetric polynomials & $1.1\pm0.03$ & this talk \\
DLCQ  & 1.38 & Harindranath \& Vary~\cite{VaryHari} \\
Quasi-sparse eigenvector & 2.5 & Lee \& Salwen~\cite{LeeSalwen} \\
Density matrix renormalization group & 2.4954(4) & Sugihara~\cite{Sugihara} \\
Lattice Monte Carlo & 
   2.70$\left\{\begin{array}{l} +0.025 \\ -0.013\end{array}\right.$ & 
   Schaich \& Loinaz~\cite{SchaichLoinaz} \\
Uniform matrix product & 2.766(5) & Milsted et al.~\cite{Milsted} \\
Renormalized Hamiltonian truncation & 2.97(14) & Rychkov \& Vitale~\cite{RychkovVitale} \\
\hline\hline
\end{tabular}
\end{table}

\section{Summary}  \label{sec:summary}

We have developed a high-order method for (1+1)-dimensional
light-front theories that is distinct from DLCQ.  The method
employs function expansions in terms of 
fully symmetric multivariate polynomials
that respect the constraint of momentum conservation.
It allows separate tuning of resolutions in each Fock sector,
and could be combined with transverse
discretization or basis functions for applications
to (3+1)-dimensional theories.

The method has been applied to $\phi^4_{1+1}$ theory, to
compute the lowest mass eigenvalues and to extract an
estimate of critical coupling for the positive $\mu^2$ case.
We have identified a systematic difference with equal-time quantization
which can be associated with the difference in mass renormalizations
of the two quantizations~\cite{Burkardt}.  We have also
compared these converged high-order Fock space truncations with the 
lowest-order LFCC calculation~\cite{LFCCphi4} and found
good agreement, which implies that the LFCC method shows
promise for rapid convergence.

\acknowledgments
This work was done in collaboration with J.R. Hiller and was
supported in part by the Minnesota Supercomputing Institute
of the University of Minnesota with grants of computing
resources.  We thank M. Burkardt and L. Martinovic for
insightful comments.

\end{document}